\documentclass[twocolumn,showpacs,preprintnumbers,amsmath,amssymb%,superscriptaddress
 aps,
 prb,
 lengthcheck,%
]{revtex4-1}
\usepackage{amssymb}
\usepackage{graphicx}% Include figure files
\usepackage{dcolumn}% Align table columns on decimal point
\usepackage{bm}% bold math
\usepackage{hyperref}% add hypertext capabilities
\usepackage{makeidx}
\makeindex
\def\>{\right\rangle}
\def\<{\left\langle}
\def\be{\begin{equation}}
\def\ee{\end{equation}}
\def\ba{\begin{array}{l}}
\def\ea{\end{array}}

\def\beq{\begin{eqnarray}}
\def\eeq{\end{eqnarray}} 

\begin{document}

\preprint{APS/123-QED}
 
\title{Composite fermions description of fractional topological insulators.}

\author{D. Ferraro$^{1,2,3}$, G. Viola$^4$}
 \affiliation{$^1$ Dipartimento di Fisica, Universit\`a di Genova,Via Dodecaneso 33, 16146, Genova, Italy.\\
$^2$ CNR-SPIN, Via Dodecaneso 33, 16146, Genova, Italy.\\ 
$^3$ INFN, Via Dodecaneso 33, 16146, Genova, Italy.\\
$^4$ Department of Condensed Matter Physics, Weizmann Institute of Science, Rehovot, 76100, Israel.}
\date{\today}% It is always \today, today,

\begin{abstract}
We propose a $\mathbb{Z}_{2}$ classification of Abelian time-reversal fractional topological insulators in terms of the composite fermions picture. We consider the standard toy model where spin up and 
down electrons are subjected to opposite magnetic fields and only electrons of the same spin interact via a repulsive force.
  By applying the composite fermions approach to this time-reversal symmetric system, we are able to obtain a hierarchy of topological insulators with spin Hall conductance $\sigma_{s}=\frac{e}{ 2\pi}\frac{p}{2mp+1} $, being $p,m \in\mathbb{N}$. They show stable edge states only for odd $p$, as a direct consequence of the Kramer's theorem.
\end{abstract}

\maketitle

Since its discovery, the Quantum Hall Effect (QHE)~\cite{DasSarma97} appeared as a totally new state of matter, outside the Ginsburg-Landau classification in terms of broken symmetries. In the Integer case, the deep connection between the quantization of the Hall conductance and the first Chern invariant was highlighted soon after the first experimental observations~\cite{TKNN}. More involved was the attempt to codify the topological features in the Fractional case~\cite{Wenbook} where the role of the electron-electron interaction is essential. An extremely fruitful way to describe the both Integer and Fractional cases in a unique framework has been provided by the concept of \emph{composite fermions} (CF), emergent excitations made up of electrons with an even number of flux quanta of the magnetic field attached~\cite{Jainbook}. By means of this cunning idea J. K. Jain has been able to describe most stable sequence of observed fractional states with filling factors $\nu=\frac{p}{2mp\pm 1}$ being $p, m \in \mathbb{N}$, as Integer QHE of CF~\cite{Jain89}. Afterwards, effective field theories in terms of coupled Abelian gauge fields have been introduced to describe states belonging to the Jain's sequence~\cite{Wen95}.  The restriction of such field theories at the boundary leads to one dimensional bosonic fields~\cite{Wen95}. 

In recent years, a new class of topological state of matter has been theoretically predicted and experimentally observed in both two and three spatial dimensions: the topological insulators (TI) (see \cite{Hasan10, Qi11} and references therein). They are characterized by a gapped bulk and gapless edge states protected by time-reversal (T) symmetry. In two spatial dimension they are realized by the Quantum Spin Hall Effect (QSHE) theoretically predicted in graphene~\cite{Kane05, Kane05b}, strained semiconductors~\cite{Bernevig06} and finally Mercury-Telluride quantum wells~\cite{Bernevig06b}, where this effect has been experimentally observed~\cite{Konig07}. A simplified description of this system has been given in terms of two Integer quantum Hall states subjected to opposite magnetic fields in order to preserve T symmetry~\cite{Bernevig06}. In real systems, the role of the spin-dependent magnetic field is played by the spin-orbit interaction. If the filling factor of each spin is odd one has helical edge states robust against T invariant perturbations and opposite spins flow in opposite directions~\cite{Kane05b}. This case represents the (Integer) QSHE. Otherwise, it is possible to gap out all the edge states without breaking T as typical of a trivial insulator. This leads to the $\mathbb{Z}_{2}$ classification of non-interacting two dimensional band insulators \cite{Kane05}. Also the case of three dimensional topological insulators have been analyzed. These systems can be  classified according to four $\mathbb{Z}_{2}$ topological invariants and they shows surface states that can be weak or strong against T invariant perturbations 
\cite{Fu07, Moore07, Ringel11}.

The observation of the Integer QSHE immediately raised interest about the possibility of fractional topological insulators in presence of strongly interacting electrons~\cite{Bernevig06}. The attention has been mainly focused on the stability of the edge states. Levin and Stern~\cite{Levin09}, assuming a toy model interaction diagonal in the electrons spin index (that conserves the third component of the spin), showed that the ratio $\sigma_{s}/e^{*}$ (in unit of $1/(2\pi)$) between the spin Hall conductance $\sigma_{s}$  and the minimal fractional charge $e^{*}$ discriminates among the trivial and topological fractional insulators. Indeed, if this ratio is even, it is possible to gap out the edge states by means of a T invariant perturbation, while in the odd case the edge states are robust and it is not possible to destroy them without breaking T explicitly or spontaneously. A more general $\mathbb{Z}_{2}$ classification of fractional topological insulators, including QSHE, have been found also for more realistic electron-electron interactions that don't necessarily conserve the projection of the spin~\cite{Kane05, Neupert11}.  A field theoretical description of the hierarchical states, T invariant extension of the Jain's sequence, has also been investigated~\cite{Cho10, Santos11}. Also the possibility of fractional TI in three dimensions have been discussed both in terms of lattice models and effective field theories \cite{Cho10, Levin11, Maciejko10}.

Aim of this paper is to provide a description of the Abelian fractional topological insulators based on the CF picture. 
 We will consider a toy model where electrons with opposite spins feel opposite magnetic field and where only electrons with the same orientation of the spin interact. 
We will operate the standard flux attachment procedure on the microscopic fermionic degree of freedom obtaining a Jain's sequence of  fractional T invariant insulators with spin Hall conductance  $\sigma_{s}=\frac{e}{ 2\pi}\frac{p}{2mp+1} $, being $p,m \in \mathbb{N}$. We will show that a natural $\mathbb{Z}_{2}$ classification emerges at the level of the bulk in terms of Integer QSHE realized by the CF. We will finally check the validity of our result by means of the Levin-Stern stability criterion.

Let us start by considering the standard toy model for fractional topological insulators~\cite{Bernevig06, Levin09}. Here, two dimensional electrons are subjected to a spin dependent static magnetic field ${\mathfrak B}=B \hat{z} \hat{\sigma}_{z}$, with magnitude $B$. With $\hat{\sigma}_{z}$ one indicates the Pauli matrix associated to the third component of the electron spin and $\hat{z}$ is the unit vector perpendicular to the electron plane. This peculiar magnetic field can be considered as a simplification of the more realistic spin-orbit interaction. In addition, a short-range two body interaction is assumed only between electrons with the same spin. The numbers of spin up and down electrons are equal.
 The above assumptions guarantee that the system is T invariant and that the third component of the spin is a good quantum number. Moreover, the system can be completely separated in the two spin sectors. For proper values of electron density and strength of interaction, one may think that the system is made up of two decoupled Fractional quantum Hall states at opposite magnetic field, \emph{i. e.} opposite Hall conductances~\cite{Levin09}.

The many-body wavefunction of the system satisfies the Schroedinger equation 
$H \Psi(\small{\textbf{r}^{(+)}_{1},..., \textbf{r}^{(+)}_{N}, \textbf{r}^{(-)}_{1}, ..., \textbf{r}^{(-)}_{N}}) =E \Psi(\small{\textbf{r}^{(+)}_{1}, ..., \textbf{r}^{(+)}_{N}, \textbf{r}^{(-)}_{1}, ..., \textbf{r}^{(-)}_{N}})$ with $E$ the energy. Due to the symmetry of the problem, it is useful to introduce the spinor notation
\be
\Psi(\small{\textbf{r}^{(+)}_{1}, ..., \textbf{r}^{(+)}_{N}, \textbf{r}^{(-)}_{1}, ..., \textbf{r}^{(-)}_{N}}) = \left( 
\ba
\Psi_{+}(\small{\textbf{r}^{(+)}_{1}, ..., \textbf{r}^{(+)}_{N}}) \\
\Psi_{-}(\small{\textbf{r}^{(-)}_{1}, ..., \textbf{r}^{(-)}_{N}})
\ea
\right)
\ee
 where, with $+$ ($-$) we indicate the quantities, as momenta and positions, associated with the $N$ electrons with spin up (down) that form the system. 
For static configurations of the external electromagnetic field, the Hamiltonian of this model can be written as
\beq
H&=&\sum_{\epsilon=\pm}\mathcal{P}^{(\epsilon)}\left\{\frac{1}{2m^{*}} \sum^{N}_{j=1} \left[ \textbf{p}^{(\epsilon)}_{j} +\epsilon\frac{e}{c} \textbf{A}(\small{\textbf{r}^{(\epsilon)}_{j}})\right]^{2} +\right. \\ 
&+&\left.\sum^{N}_{i<j} V(\small{\textbf{r}^{(\epsilon)}_{i}}-\small{\textbf{r}^{(\epsilon)}_{j}})+\sum^{N}_{j=1}\left(eA_{0}(\small{\textbf{r}^{(\epsilon)}_{i}})-\mu\epsilon B(\small{\textbf{r}^{(\epsilon)}_{i}})\right)\right\}
\nonumber
\label{H_0}
\eeq
 with $m^{*}$ the electron effective mass, $-e$ the electron charge, $c$ the speed of light, $\mu$ half of the electron magnetic momenta in the material. 
 The projectors
\be
\mathcal{P}^{(\pm)}=\frac{1}{2} \left( \hat{I}\pm \hat{\sigma}_{z}\right),
\ee
with $\hat{I}$ the identity, act on the spin space.
The external vector potential $\textbf{A}$ satisfies $\nabla \times \textbf{A}=B \hat{z}$. It is easy to note that only the magnetic field couples with opposite signs with electrons of different spins as it is indicated by the minimal coupling in the Hamiltonian. Since the $\hat{z}$ component of the spin is conserved, the last term of the Hamiltonian reduces to an additive constant and will be neglected in the following. From now on the scalar potential $A_{0}$ will be fixed to be zero. We don't need to enter into the details of the form of the two body interaction $V$, that however can be thought as a screened Coulomb potential.
 
The complete solution of the many-body Hamiltonian in Eq. (\ref{H_0}) is an hopeless task, nevertheless it is possible to properly manage the problem in order to treat it at the mean field level~\cite{Simon99, Jainbook}.

Firstly, it is convenient to make the so called Chern-Simons transformation 
\be
\Phi_{\pm}= \left[ \prod_{i<j} e^{\mp i2m \theta(\textbf{r}^{(\pm)}_{i}- \textbf{r}^{(\pm)}_{j})}\right] \Psi _{\pm}
\label{transformation}
\ee
that acts on the phases of the two components of the many-body wavefunction. The multiple-valued function $\theta(\small{\textbf{r}^{(\pm)}_{i}}-\small{\textbf{r}^{(\pm)}_{j}})$ represents the angle formed by the vector $\small{\textbf{r}^{(\pm)}_{i}}-\small{\textbf{r}^{(\pm)}_{j}}$ with an arbitrary $\hat{x}$ axis in the plane containing the electrons.

It is easy to note that the considered transformation does not modify the fermionic nature of the excitations for $m\in \mathbb{N}$ and that the transformed wavefunction 
\be
\Phi(\small{\textbf{r}^{(+)}_{1}, ..., \textbf{r}^{(+)}_{N}, \textbf{r}^{(-)}_{1}, ..., \textbf{r}^{(-)}_{N}})  = \left( 
\ba
\Phi_{+}(\small{\textbf{r}^{(+)}_{1}, ..., \textbf{r}^{(+)}_{N}}) \\
\Phi_{-}(\small{\textbf{r}^{(-)}_{1}, ..., \textbf{r}^{(-)}_{N}})
\ea
\right)
\ee
 is a solution of the Schroedinger equation $H_{\mathrm{CF}} \Phi(\small{\textbf{r}^{(+)}_{1}, ..., \textbf{r}^{(+)}_{N}, \textbf{r}^{(-)}_{1}, ..., \textbf{r}^{(-)}_{N}}) =E \Phi(\small{\textbf{r}^{(+)}_{1}, ..., \textbf{r}^{(+)}_{N}, \textbf{r}^{(-)}_{1}, ..., \textbf{r}^{(-)}_{N}})$ where the new Hamiltonian reads
\beq
H_{\mathrm{CF}}&&=\sum_{\epsilon=\pm}\mathcal{P}^{(\epsilon)}\left\{\frac{1}{2m^{*}} \sum^{N}_{j=1} \left[ \textbf{p}^{(\epsilon)}_{j} +\epsilon\frac{e}{c}[ \textbf{A}(\small{\textbf{r}^{(\epsilon)}_{j}})-\textbf{a}(\small{\textbf{r}^{(\epsilon)}_{j}})]\right]^{2} \right.\nonumber \\ 
&&+\left.\sum^{N}_{i<j} V(\small{\textbf{r}^{(\epsilon)}_{i}}-\small{\textbf{r}^{(\epsilon)}_{j}})\right\}
\eeq
being
\be
\textbf{a} (\textbf{r}^{(\pm)}_{i})=\frac{2 m \phi_{0}} {2 \pi} \sum^{N}_{j=1} \frac{\hat{z} \times (\small{\textbf{r}^{(\pm)}_{i}}-\small{\textbf{r}^{(\pm)}_{j}})}{|\small{\textbf{r}^{(\pm)}_{i}}-\small{\textbf{r}^{(\pm)}_{j}}|^{2}}
\ee
a Chern-Simons vector potential. With $\phi_{0}=hc/e$ we indicate the elementary magnetic flux quantum.

Due to the fact that the function $\theta$ is not singe-valued, $\nabla \times \textbf{a} (\textbf{r}) =2 m \phi_{0} \delta(\textbf{r}- \textbf{r}^{(\pm)}_{j})$ is singular in correspondence of the positions of the electrons. This leads to an additional spin dependent Chern-Simons magnetic field 
\be
\mathfrak{ b}(\textbf{r}) = 2 m\phi_{0}\hat{\sigma}_z  \left( 
\ba
n^{(+)}(\textbf{r})\\
n^{(-)}(\textbf{r})
\ea
\right)
\ee 
with $n^{(\pm)}(\textbf{r})=\sum^{N}_{j=1}\delta(\textbf{r}-\textbf{r}^{(\pm)}_{j})$ the density operator for the electrons with spin up and spin down respectively. The procedure we have carried out leads to the attachment of $2m$ quanta of magnetic field flux to each electron, whose sign depends on the spin orientation.

The transformation in Eq. (\ref{transformation}) only acts as a spin dependent phase factor for the electronic wavefunction and does not modify its amplitude. Therefore the density operators $n^{(\pm)}$ are the proper one also for the transformed fermions, that in the following we will indicate as CF \cite{Jainbook, Jain89, Simon99}. 

At the mean field level one assumes a uniform density for the electrons, and consequently for the CF.  The resulting magnitude of the Chern-Simons magnetic field is
\be
\langle b \rangle=  2 m \phi_{0} n
\label{flux_attach}
\ee
being $n=\langle n^{(+)}\rangle=\langle n^{(-)}\rangle$ the average two dimensional electron density equal for both spins. 

According to definition in Eq. (\ref{transformation}) the effective Chern-Simons field is opposite to the physical magnetic field $B$ for both spins, therefore the CF in the system feel a reduced magnetic field of magnitude
\be
\tilde{B}=B- \langle b \rangle
\label{reduced_B}
\ee
with opposite direction for the two spins. Moreover, due to the assumption of uniform density, the electron-electron interaction reduces to a constant and does not affect anymore the dynamics of the system.

Through the previous steps we have been able to map the initial problem of interacting electrons in a strong spin dependent magnetic field into the one of free CF subjected to a reduced magnetic field with direction driven by the spin orientation. This new system is described in terms of the mean field Hamiltonian 
\be
H_{mf}=\sum_{\epsilon=\pm}\mathcal{P}^{(\epsilon)}\left\{\frac{1}{2m^{*}} \sum^{N}_{j=1} \left[ \textbf{p}^{(\epsilon)}_{j} +\epsilon\frac{e}{c}\tilde{\textbf{A}}(\small{\textbf{r}^{(\epsilon)}_{j}})\right]^{2}\right\}
\label{Hmean}
\ee
with $\tilde{B}= \nabla \times \tilde{\textbf{A}}$ directed along $\hat{z}$.

One can define now the composite fermions filling factors  $\tilde{\nu}^{(\pm)}=\pm n \phi_{0}/\tilde{B}$ for each spin. In analogy with what observed in the QHE it is reasonable to assume that each spin has states at $\tilde{\nu}^{(\pm)}=\pm p$ with $p\in \mathbb{N}$ that remain stable also when perturbative correction to the mean field  approximation are taken into account \cite{Simon99}. For sake of simplicity we only focused on the positive $p$ case. For negative value of $p$, one is in presence of CF created by holes instead of electrons, but all the previous considerations hold as well.

Spin up and down CF form two Integer QHE with opposite direction of the magnetic field. This T invariant system can be seen as an Integer QSHE of CF \cite{Kane05, Bernevig06, Bernevig06b}. Recalling Eqs. (\ref{flux_attach}) and (\ref{reduced_B}), the correspondent electron filling factors are 
\be
\nu^{(\pm)}=\pm \frac{p}{2mp+1}
\ee
according to the formation of a Fractional QSHE of states in the Jain sequence supporting a spin Hall conductance $\sigma_{s}=\frac{e}{ 2\pi}\frac{p}{2mp+1} $ \cite{Neupert11, Santos11}.

An important question to be faced at this stage is the robustness of the edge states of the considered system against disorder and other possible T invariant perturbations. Insulators can be divided into two distinct classes: trivial and topological~\cite{Kane05}. In the simple case of a T invariant systems with conserved third component of the spin one has $\nu^{(+)}+\nu^{(-)}=0$, namely a null first Chern number (vanishing Hall conductance)~\cite{TKNN}. Nevertheless, it is possible to define the additional topological quantum number
\be
\eta= \frac{1}{2} \left(\nu^{(+)}-\nu^{(-)}\right)\qquad \mathrm{mod}~(2).
\ee
For $\eta=0$ one has a trivial insulator equivalent to the vacuum, while for $\eta=1$ one has a topological insulator~\cite{Kane05}. This $\mathbb{Z}_{2}$ classification is a consequence of the Kramer theorem for a T invariant system of half-integer spin electrons. As a remarkable outcome, the boundary of an Integer QSHE has edge states robust against T invariant interaction terms only for odd filling factors of each spin $(\eta=1)$, that manifests as an odd number of pair of counter-propagating edge channels related by T (Kramer's doublets). In other words, for a T invariant electron system with an odd number of Kramer's doublets  is not possible to gap out all the the edge states without breaking T. Otherwise when the number of pairs is even a T invariant perturbation is able to induce a gap for the states at the boundary. 

The above general argument based on the topological properties of the bulk material and the Kramer theorem can be directly applied to the system of non-interacting CF described by the mean field Hamiltonian in Eq.~\ref{Hmean} and results in the existence of robust edge states only for $p$ odd integer. 
Obviously this condition on $p$ holds as well for the original electron system that inherits the topological classification from the CF system. Therefore, in terms of the sorting introduced in Ref.~\cite{Levin09}, one has ``fractional topological insulators" only for odd $p$ and ``fractional trivial insulators" otherwise.  The previous discussion clearly shows that the CF picture reveals extremely useful in order to extract relevant physical information about the stability of the states in a simple way also for fractional TI, exactly as in the case of the FQHE. As a final remark we need to underline that the presented approach only works for the hierarchical states where the CF picture is well established. The analysis of states like $\nu^{(\pm)}=\pm 1/2$, that could be though as a T invariant generalization of the Pfaffian model introduced for the Fractional QHE~\cite{Moore91} is out of the aim of this paper. In the following we will consider the Levin-Stern criterion for the stability of the edge states~\cite{Levin09} for topological insulator in Jain's sequence, showing its equivalence with the $\mathbb{Z}_{2}$ classification of the bulk states discussed above in terms of the CF picture.

We consider now a low energy effective description able to enclose the properties of T symmetric fractional states introduced above. In the framework of edge effective field theory for the Fractional QHE, the states belonging to the Jain's sequence are described in terms of the Lagrangian density \cite{Wen95}
\beq
\mathcal{L}_{c}(\varphi, K, V, t)&&= \frac{1}{4 \pi} \left( K^{i j} \partial_{x} \varphi_{i} \partial_{t} \varphi_{j} -V^{ij} \partial_{x} \varphi_{i} \partial_{x} \varphi_{j} +\right.\nonumber \\
&&\left.
+\varepsilon^{\mu \nu} t^{i} \partial_{\mu} \varphi_{i} A_{\nu}\right)
\eeq
being $\varphi$ a $p$-component vector of bosonic fields, $A_{\mu} $ the external vector potential and $\varepsilon$ the total antisymmetric tensor in two dimensions. The universal $p\times p$ symmetric matrix $K$ in the so called symmetric basis is $K _{i,j}=\delta_{i,j}+2m  C_{i,j}$ with $C_{i,j}=1\,\,\forall \,\, i,~j$, it encloses the topological order of the states. The $p$-component vector $t=(1,...,1)$ couples the bosons with the external electromagnetic field. 
 The non universal $p\times p$ symmetric matrix $V$ contains interaction effects at the boundary. Greek indices denote the time and space coordinates while the Latin ones label the $p$ bosonic fields. Sums over the repeated indices are underlined.

The simplest T invariant extension of the above model is give by \cite{Levin09}
\be
\mathcal{L}=\mathcal{L}_{c}(\varphi_{+}, K, V, t)+\mathcal{L}_{c}(\varphi_{-}, -K, V, t).
\label{Lagrangian}
\ee
It is worth to note that the above Lagrangian density for the edge states can be seen as a boundary restriction of the double Chern-Simons or equivalently of the BF theory introduced to model the bulk of the T invariant system
~\cite{Santos11, Cho10, Blasi12}. It describes $p$ couples of counter-propagating edge states connected by T symmetry, namely $p$ Kramer doublets. A more general definition which also admits the coupling between the $\varphi_{+}$ and $\varphi_{-}$ bosons is possible \cite{Santos11}, however its discussion is out of the aim of this paper.

Levin and Stern~\cite{Levin09} introduced a criterion to indicate whenever the edge structure of Abelian fractional T invariant states of matter is robust against perturbations that don't break T symmetry. By means of a general flux insertion argument~\cite{Fu06}, further supported by explicit effective field theory considerations, they proved that the stability depends on the parity of the ratio between the spin Hall conductance $\sigma_{s}$ (in unit of $e/2\pi$) and the minimal fractional charge predicted by the model $e^{*}$ (in unit of $e$).  If this ratio is even, a T invariant perturbation is able to gap out the edge states, otherwise it is not possible to localize them without breaking T symmetry.

Starting from Eq. (\ref{Lagrangian}) we can define the $2 p \times 2p$ matrix 
\be
\tilde{K}=\left( 
\begin{array}{cc}
K & 0\\
0 & -K \\
\end{array}
\right)
\ee
and the vector $\tilde{t}=(t,t)$, T invariant extension of $t$. Moreover, due to the additional $U(1)$ symmetry related to the conservation of the spin projection, it is possible to introduce the vector $\tilde{s}=(t/2, -t/2)$ that couples the bosonic field with an external fictitious Abelian gauge field \cite{Santos11}. The coupling with localized elementary quasiparticle excitations, namely the ones with minimal charge, involve the $2p$-components vector $\tilde{l}=(1, 0,...,0)$. Due to the peculiar form of the matrix $\tilde{K}$, any vector with one entry equal $1$ in the first $p$ slots or to $-1$ in the last $p$ slots and the other null provide the same description of the elementary excitations that are therefore $2p$ times degenerate.
  
It is easy to note that \cite{Neupert11}
\beq
\sigma_{s}&=& \tilde{s}_{i} (\tilde{K}^{-1}) ^{ij} \tilde{t}_{j}\frac{e}{  2\pi} =\frac{p}{2mp+1}\frac{e}{  2\pi}
\label{sigma}\\
e^{*}&=& \tilde{l}_{i} (\tilde{K}^{-1}) ^{ij} \tilde{t}_{j}e =\frac{1}{2mp+1} e
\label{e_star}
\eeq
and that the Hall conductance of the system $\sigma_{c}= \tilde{t}_{i} (\tilde{K}^{-1}) ^{ij} \tilde{t}_{j}\frac{e^{2}}{ h}$ is zero as required by T symmetry.

By comparing Eqs. (\ref{sigma}) and (\ref{e_star}) one has that the ratio $\sigma_{s}/ e^{*}$ (in units of $1/2\pi$) is equal to $p$. Therefore, for odd $p$ the edge structure is robust against T invariant perturbations, while for even $p$ the states at the boundary can be gapped out without breaking T. This scenario is in complete agreement with the one we have previously derived in terms of the CF picture.

To conclude, in this paper we proposed a description of Abelian fractional topological insulators in the language of the CF that had already revealed extremely fruitful in the framework of the Fractional QHE. Our approach, valid for the Jain's states supporting spin Hall conductance $\sigma_{s}=\frac{e}{ 2\pi}\frac{p}{2mp+1} $, represents a natural way to introduce a $\mathbb{Z}_{2}$ classification also for fractional topological insulators at the level of the bulk. We observed that for odd $p$ the edges of this peculiar states of matter are robust due to the Kramers theorem applied to the CF and form ``fractional topological insulators", while for even $p$ edges can be gapped out without breaking T symmetry leading to ``fractional trivial insulators". This result is in agreement with the Levin-Stern stability criterion. The presence of spin-mixing interaction could be perturbatively taken into account starting from the presented CF  approach. The extension of this idea to states out of the Jain sequence and supporting non-Abelian excitations like $\nu^{(\pm)}=\frac{1}{2}$ will be an important goal for future developments.

\acknowledgments
 We thank M. Sassetti, N. Magnoli, A. Braggio, M. Carrega, M. Koch-Janusz, Jiang J. Hua and A. Stern for valuable discussions.  D. F.  acknowledges the support of the CNR STM 2010 programme and the EU-FP7 via ITN-2008-234970 NANOCTM, G. V. acknowledges ``Feinberg Graduate School" and ``Fondazione della Riccia" for the financial support.

\end{document}